\documentclass[twocolumn,showpacs,pra,amsmath,amssymb,superscriptaddress,amssymb]{revtex4}

\usepackage{graphicx}
\usepackage{dcolumn}
\usepackage{bm}
\usepackage[dvipdfm]{xcolor}
\usepackage{subfigure}
\usepackage{color}

\begin{document}

\title{Exploring multi-band excitations of interacting Bose gases in a 1D optical lattice by coherent scattering}

\author{Xinxing Liu}
\affiliation{School of Electronics Engineering and Computer Science,
Peking University, Beijing 100871, P.R. China}
\author{Xiaoji Zhou}\thanks{Electronic address: xjzhou@pku.edu.cn}
\affiliation{School of Electronics Engineering and Computer Science,
Peking University, Beijing 100871, P.R. China}
\author{Wei Zhang}
\affiliation{Department of Physics, Renmin University of China, Beijing 100872, P.R. China}
\author{Thibault Vogt}
\affiliation{School of Electronics Engineering and Computer Science,
Peking University, Beijing 100871, P.R. China}
\author{Bo Lu}
\affiliation{School of Electronics Engineering and Computer Science,
Peking University, Beijing 100871, P.R. China}
\author{Xuguang Yue}
\affiliation{School of Electronics Engineering and Computer Science,
Peking University, Beijing 100871, P.R. China}
\author{Xuzong Chen}
\affiliation{School of Electronics Engineering and Computer Science,
Peking University, Beijing 100871, P.R. China}

\date{\today}

\begin{abstract}
We use a coherent Bragg diffraction method to impart an external
momentum to ultracold bosonic atoms trapped in a one-dimensional optical
lattice. This method is based on the application of a single light pulse, with conditions where scattering of photons can be resonantly amplified by the atomic
density grating. An oscillatory behavior of the momentum
distribution resulting from the time evolution in the lattice
potential is then observed. By measuring the oscillating frequencies, we
extract multi-band energy structures of single-particle excitations
with zero pseudo-momentum transfer for a wide range of lattice
depths. The excitation energy structures reveal the interaction
effect through the whole range of lattice depth.
\end{abstract}

\maketitle

\section{introduction}
\label{sec:introduction}

Cold atomic gases loaded in optical lattices have drawn great attention in the
past few years as a versatile and powerful experimental system to study
strongly correlated many-body problems~\cite{bloch-05}. By taking advantage of
great controllability and perfectness, this novel type of matter features a
promising candidate to achieve some important modeling systems and
help tackling many unaddressed questions in multi-disciplinary fields of physics.
As an example, important progress has been made in manipulation and
characterization of interacting bosons loaded in an optical lattice, both for the
low-lattice/low-temperature Bose-Einstein condensate (BEC)
regime and for the high-lattice/high-temperature Mott Insulator (MI) or
normal regime~\cite{greiner-02, stoferle-04, paredes-04, kinoshita-04,
chin-06, hadzibabic-06, spielman-07}.

Among these efforts, the understanding of excitations gives crucial information
about the underlying system, for its ability to characterize the system's response
in different regimes. For shallow lattices, the gas behaves as superfluid
and the excitation spectrum can be described by the Bogoliubov
theory within a mean-field description of interaction effect. For deep lattices,
the gas enters the strongly correlated MI phase, where the excitation exhibits
a gap at low energies. Experimental characterization of these features include
the measurement of first band effective mass by Bloch
oscillation~\cite{morsch-01}, the demonstration of gap in the MI regime~\cite{stoferle-04},
and the momentum-resolved probe of multi-band dispersion by Bragg
spectroscopy~\cite{kozuma-99,stenger-99,papp-08,veeravalli-08,clement-09,ernst-10}.

\begin{figure}[t]
\centering
\includegraphics[width=8cm]{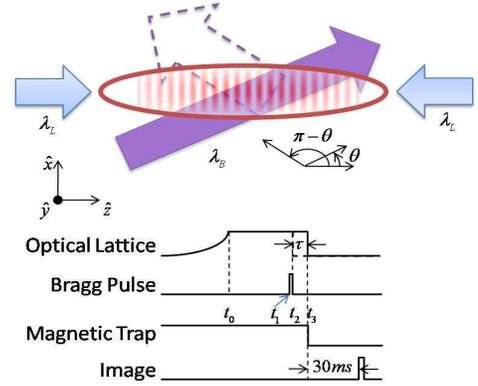}
\caption{(Color online)Schematic setup (top) and time sequence (bottom) of
our experimental scheme. After loading the BEC in a 1D optical lattice
with an exponential ramp (40 ms), a laser beam of $5 \mu$s is sent at $\theta=24^{\circ}$
from the longitudinal axis, and is scattered into a symmetrical direction
according to the Bragg scattering condition. After holding the lattice potential for
another duration of $\tau$, both the lattice and the magnetic trap are abruptly switched
off and an absorption image is taken after a time-of-flight (TOF) of 30 ms.}
\end{figure}

In this manuscript, we report a novel experimental scheme to detect
multi-band excitation with zero quasi-momentum transfer for
interacting bosons loaded in a one-dimensional (1D) optical lattice
across a wide range of lattice depth. The elementary excitations are
created using Bragg scattering, where light waves are scattered
collectively in accommodation to the periodic distribution of
bosonic atoms. As a response to this coherent process, the atoms
will recoil with a momentum in unit of the reciprocal lattice, and
can be excited to higher energy bands with the same pseudo-momentum.
By monitoring the time evolution of the momentum distribution, we
can extract information about the excitation energy. In shallow
lattices where gases are superfluid, we can get the $(n=1)$ as well
as the $(n=2)$ band excitation energy. For deep lattices where the
ground state is a MI, the large band gap prohibits dramatic
population to higher band, such that only the $(n=1)$ band
excitation energy is obtained. Throughout the whole range of lattice
depth variation, we also observe a sizable signature of interaction
effect, due to the large occupation number for each lattice site.

We stress that our scheme is different from Bragg spectroscopy,
where atoms are excited by a direct transfer of energy and momentum
via a two-photon process stimulated by two distinct laser
beams~\cite{kozuma-99,stenger-99,papp-08,veeravalli-08,clement-09,ernst-10}.
In Bragg spectroscopy, the two laser beams form an effective
one-dimensional optical lattice moving along a specified direction.
If we see from the lattice frame, the atoms which are stationary in
the lab frame are moving along the opposite direction, and the Bragg
spectroscopy scheme corresponds to an effective diffraction of
atomic wave function on this optical grating. Therefore, this
procedure usually requires a coherence condition for atoms and a
beam duration of milliseconds to achieve distinct diffraction peaks,
which corresponds to significant population of specific excited
states.

In contrast, our experimental scheme involves only one incident
beam, and the periodic density distribution of atoms in optical
lattices are considered as gratings, from which the photons are
scattered and cooperatively amplified given a certain geometry to
fulfill the Bragg scattering condition. Therefore, this method does
not rely on phase coherence of atoms in different sites, and can be
directly applied to the MI/normal regime. Besides, as the speed of
photon is $c$, the building up time of scattering light field is
negligible and this optically coherent process requires a much
shorter light beam duration only in scale of microseconds, hence
brings less heating or other complications to the atomic system.

Our paper is organized as follows. In Sec. \ref{sec:experiment} we
describe the experimental scheme based on the application of a single Bragg diffraction pulse on a gas loaded in a one dimensional lattice. In Sec. \ref{sec:result} we present our
results on the oscillatory behavior of the momentum distribution observed as a function of keep time in the optical lattice after applying the Bragg pulse. Sec.
\ref{sec:theory} is devoted to a more careful analysis of Bragg diffraction obtained with a single light pulse and used as a tool throughout this experiment. This diffraction is revisited with semi-classical theory and an interpretation in terms of collective light scattering; In Sec. \ref{sec:conclusion}, we summarize our
work and draw conclusion.



\section{Experimental description}
\label{sec:experiment}

The experimental setup and protocol is illustrated in Fig.1. We
first prepare a BEC of about $2\times 10^{5}$ $^{87}$Rb atoms in a
QUIC trap, with longitudinal length $L =100$ $\mu$m and transverse
length $l =10$ $\mu$m. The BEC is then loaded into a 1D optical
lattice $V_{\rm ol}(x) = V [1+ \cos(2 k_L x)]/2$ along its axial
direction by an adiabatic ramping of $40$ ms. Here, $V$ is the
lattice depth, and $ k_L = 2 \pi / \lambda_L$ is the reciprocal
lattice spacing with $\lambda_L$ the lattice light wavelength. After
holding the lattice for $50$ ms, we shine the Bragg pulse from a
certain direction $\theta$ (see Fig. 1). The beam is red-detuned by
$\delta=1.3$ GHz from the $ (5 S_{1/2}, F=2) \rightarrow (5 P_{3/2},
F^{'} =3)$ transition, and for a typical duration of  $\Delta
t=t_2-t_1=5$ $\mu$s. After the Bragg pulse, we further hold the
system for a certain amount of delay time $\tau$, and suddenly shut
down all the trapping potentials to get time-of-flight (TOF) images
which are taken after 30 ms of free expansion process.

Due to the presence of the optical lattice, the atoms are initially
distributed into an array of quasi-two-dimensional slices, which are
separated equidistantly by the lattice spacing $\lambda_L/2 = 426$
nm. For each slice of atoms, the Bragg pulse incident from the
direction of $\theta$ will be {\it reflected} to the direction of
$(\pi - \theta)$, such that the optical lengths of different light
paths are the same. Besides, the coherent condition for diffraction
also requires the optical length difference between light paths
scattered from adjacent slices is a multiple of the Bragg beam
wavelength $\lambda_B$. In our case of $\lambda_B = 780$ nm, these
coherent conditions single out a specific incident direction of the
Bragg pulse with $\cos \theta = \lambda_B/\lambda_L$, only at which
significant collective scattering of light could occur.

\begin{figure}[t]
\centering
\includegraphics[width=4.3cm]{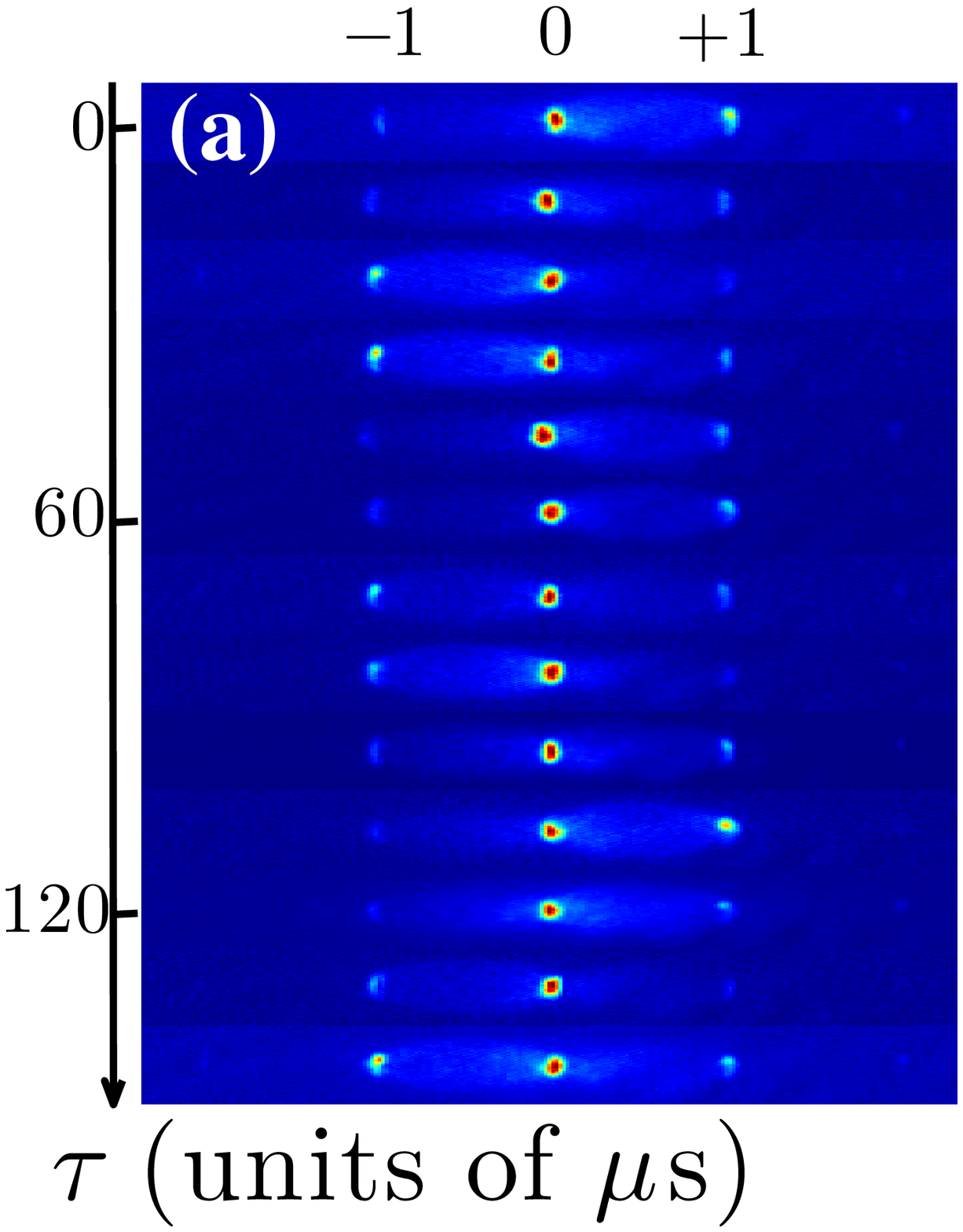}
\hspace{-0.5cm}
\includegraphics[width=4.3cm]{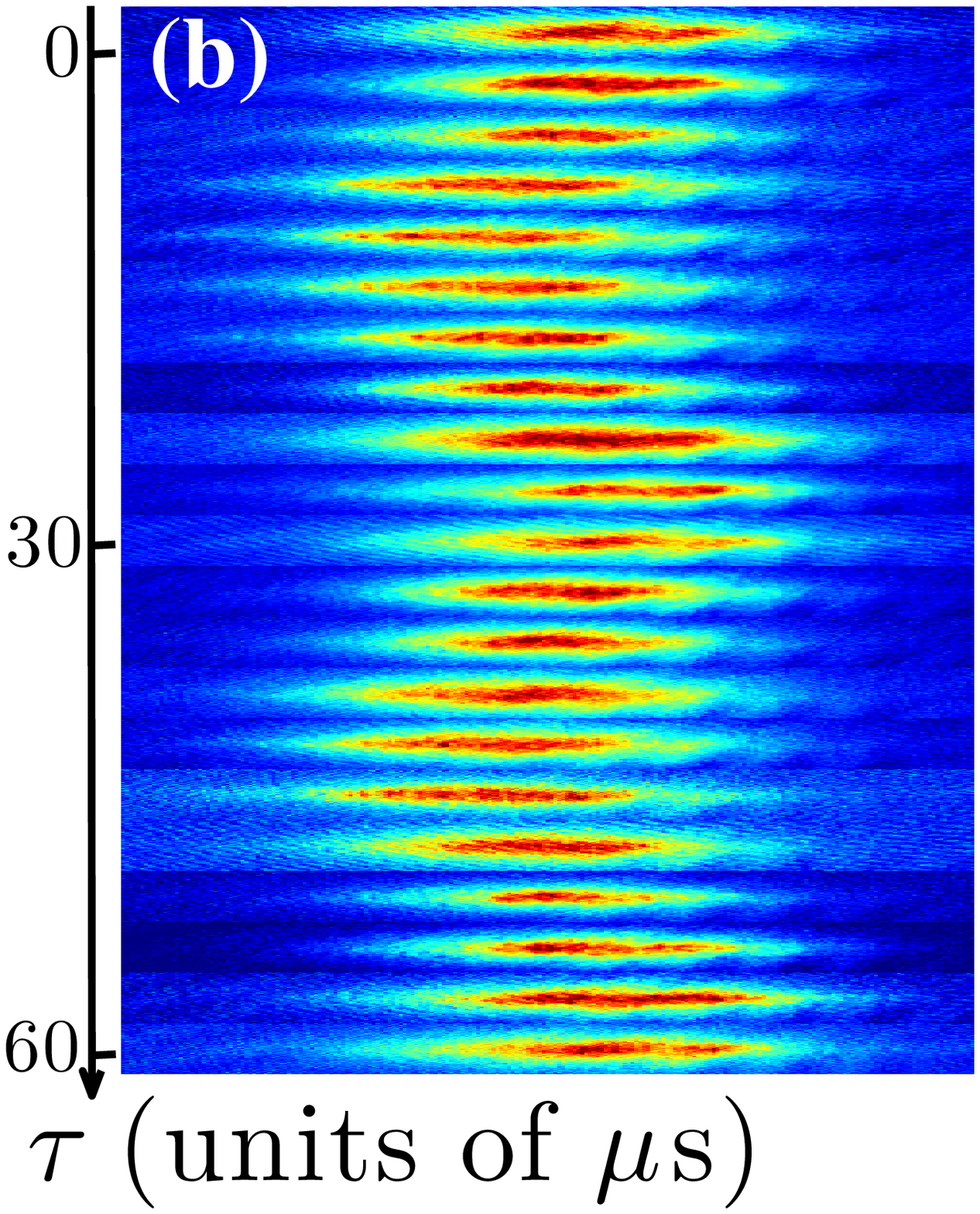}
\caption{(Color online) (a) TOF images for $V=12E_R$. From top to
bottom, the holding time $\tau$ increases from zero with a step of
$12\mu$s. An oscillatory behavior of the first order peaks at $\pm 2
\hbar k_L$ can be clearly observed. (b) Similar results for $V=37
E_R$. Although the interference pattern is wiped out and no peaks is
present, the oscillation of the momentum distribution can be still
determined.}
\end{figure}

As a consequence of cooperative scattering (cf. Sec. \ref{sec:theory}), the photons would exert
a recoil momentum of $2\hbar k_L$ to the atoms, leading to the
occupation of different states with various bands but zero quasi
momentum. After that, the momentum distribution experience an
obvious oscillation due to the presence of optical lattices. This
oscillation can be observed from a series TOF images by varying the
holding time $\tau$. One typical result for $V = 12 E_{\rm R}$ is
shown in Fig. 2(a), where the system is in the BEC regime and a
clear oscillation of the occupation number of particles in $p = \pm
2 \hbar k_L$ is observed. In Fig. 2(b), we also present a similar
result for $V = 37 E_{\rm R}$, where the high lattice potential
destroys the phase coherence and the system becomes a normal gas.
Notice that although the interference patterns no longer exist in
the TOF images, the momentum oscillation can still be clearly
determined.


\section{Results and analysis}
\label{sec:result}

For shallow lattices where the system is in the BEC phase, atoms are
macroscopically distributed in the lowest quantum state which is
described by the Gross-Pitaevskii (GP) equation
\begin{equation}
\label{eqn:GP}
\left[ -\frac{\hbar^2\partial_x^2 }{2m} + V_{\rm ol}(x) + V_{\rm ext}(x)
+ U_0 |\Psi(x)|^2 \right] \Psi(x) = \mu \Psi(x),
\end{equation}
where $V_{\rm ext}(x) = m\omega_x^2 x^2 /2$ is the external harmonic
trap, $U_0 = 4 \pi \hbar^2 a_s /m$ is the interaction with $a_s$ the
$s$-wave scattering length, and $\mu$ is the chemical potential. In
the mean field level, the repulsive interaction acts as an effective
potential $U_0 n_0(x)$ with $n_0$ the condensate fraction, which
tends to reduce the height of the optical lattice. This combined
effective lattice $V_{\rm eff}(x) = V_{\rm ol}(x) + U_0 n_0(x)$ has
the same period as the original one, and forms a band structure for
single particle state. The presence of BEC ensures that only the
state at the band bottom is macroscopically occupied. This state is
characterized by the band index $(n=0)$ and pseudo-momentum $(q =
0)$, denoted by
\begin{equation}
\label{eqn:BECstate}
\Psi(x, t_1) = \phi_{n=0, q=0} (x) = \sum_R W_{n=0}(x-R),
\end{equation}
where $\phi_{n,q}(x)$ is the Bloch state, $W_n(r)$ is the Wannier function,
and the summation over $R$ runs over all lattice sites. After the Bragg
pulse, an external momentum of $2 \hbar k_L$ is transferred to the
atoms. As this momentum transfer is resonant with the reciprocal
lattice, there should be no new pseudo-mementum state being excited,
and the system is described by
\begin{eqnarray}
\label{eqn:new state}
\Psi(x, t_2) &=& \sum_{n} a_n \phi_{n, q=0} (x)
\nonumber \\
&=& L^{-1/2} \sum_{n,\ell} a_n c_{n, q = 0}^{\ell}
e^{i 2 \ell \hbar k_L x},
\end{eqnarray}
where $c_{n, q}^{\ell}=\langle 2 {\ell}\hbar k_L|\phi_{n, q} \rangle$
is the projection of the Bloch state to the plane wave state with
momentum $2 {\ell}\hbar k_L$,
and $a_n$ is determined by the intensity and duration of the Bragg beam.
Here, we should emphasize that the Bloch states $\phi_{n,q}(x)$ with
$q=0$ are periodic functions of $\lambda_L/2$, hence can be further
expanded by a series of plane waves with $\ell$ being integers.

After the Bragg pulse, the wavefunction $\Psi(x,t)$ evolves in the
presence of the optical lattice as
\begin{eqnarray}
\label{eqn:state evolve}
\Psi(x,t_2+\tau) &=& \sum_n a_n \phi_n (x) e^{i E_n \tau}
\nonumber \\
&=& L^{-1/2} \sum_\ell e^{i 2 \ell \hbar k_L x}
\sum_n a_n c_n^\ell e^{i E_n \tau}.
\end{eqnarray}
Here, we have dropped the subscript $(q=0)$ for simplify notation,
and $E_n$ is the corresponding eigenenergy of the $n^{\rm th}$ band.
In realistic conditions, the summation over $n$ is restricted to $n
\le 2$ due to negligible population of higher bands. Therefore, the
average momentum $\langle p \rangle=\langle\Psi(t)|p|\Psi(t)\rangle$
and the atom number in the $\ell ^{th}$ momentum mode $N_{\ell}$
would acquire oscillatory behavior
\begin{eqnarray}
\frac{\langle p \rangle}{2 \hbar k_L} &=& C_{10} \cos(\omega_{10}
\tau+\varphi_{10})
\nonumber \\
&&  \hspace{1cm} + C_{21} \cos(\omega_{21} \tau+\varphi_{21}),
\label{eqn-average p}\\
\frac{N_\ell-\overline{N_\ell}}{N_{\rm tot}} &=& D_{10}^\ell
\cos(\omega_{10} \tau+\varphi_{10}) + D_{20}^\ell \cos(\omega_{20}
\tau+\varphi_{20})
\nonumber \\
&& \hspace{1cm} + D_{21}^\ell \cos(\omega_{21} \tau+\varphi_{21}),
\label{eqn:occupation}
\end{eqnarray}
where the three frequencies correspond to the energy gaps of
$\omega_{10} = (E_1 - E_0)/\hbar$, $\omega_{20} = (E_2 -
E_0)/\hbar$, and $\omega_{21} = (E_2 - E_1)/\hbar$, respectively, as
illustrated in Fig. 3(c). Notice that the frequency  $\omega_{20} =
(E_2 - E_0)/\hbar $ is not present in the oscillation of $\langle
p\rangle$. This is because the momentum distribution of states in
bands $(n=0)$ and $(n=2)$ are both even symmetric, hence make no net
contribution to $\langle p \rangle$. The coefficients $C$'s and
$D$'s are determined by
\begin{eqnarray}
\label{eqn:coeffients}
C_{nm} &=& 2|a_m^{\ast}a_n|\sum_\ell \ell c_m^{\ell}c_n^{\ell},
\nonumber \\
D_{nm}^{\ell} &=& 2|a_m^{\ast}a_n| c_m^{\ell}c_n^{\ell},
\end{eqnarray}
the initial phase factors are determined by $\varphi_{nm}={\rm
angle}(a_m^{\ast}a_n)$, the $N_{\rm tot}$ is the total atom number
and the time average of $N_\ell$ is
$\overline{N_\ell}=\sum_n|a_n|^2|c_n^{\ell}|^2 N_{\rm tot}$.

\begin{figure}[t]
\centering
\includegraphics[width=8.5cm]{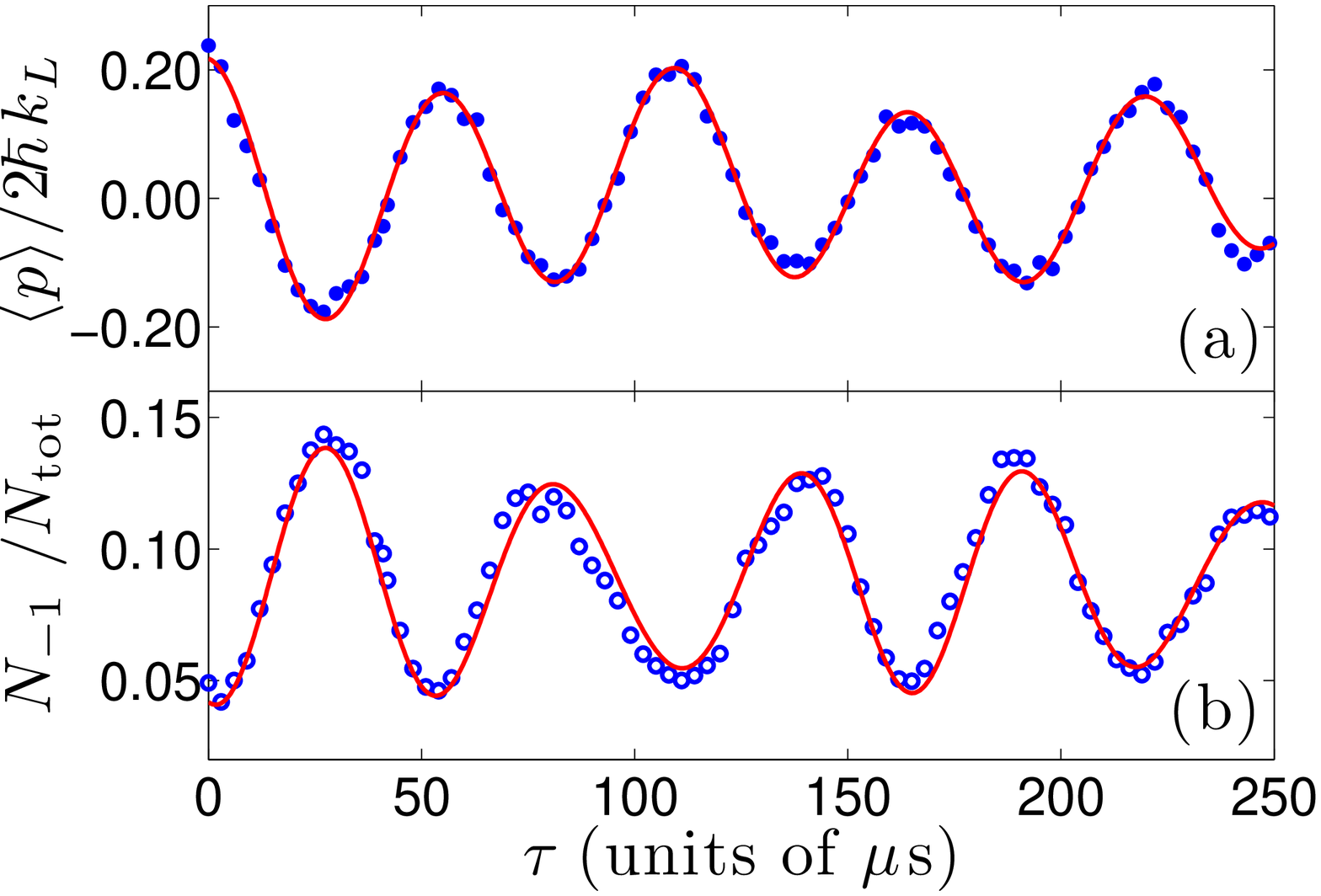}
\\
\hspace{2mm}
\includegraphics[width=8.3cm]{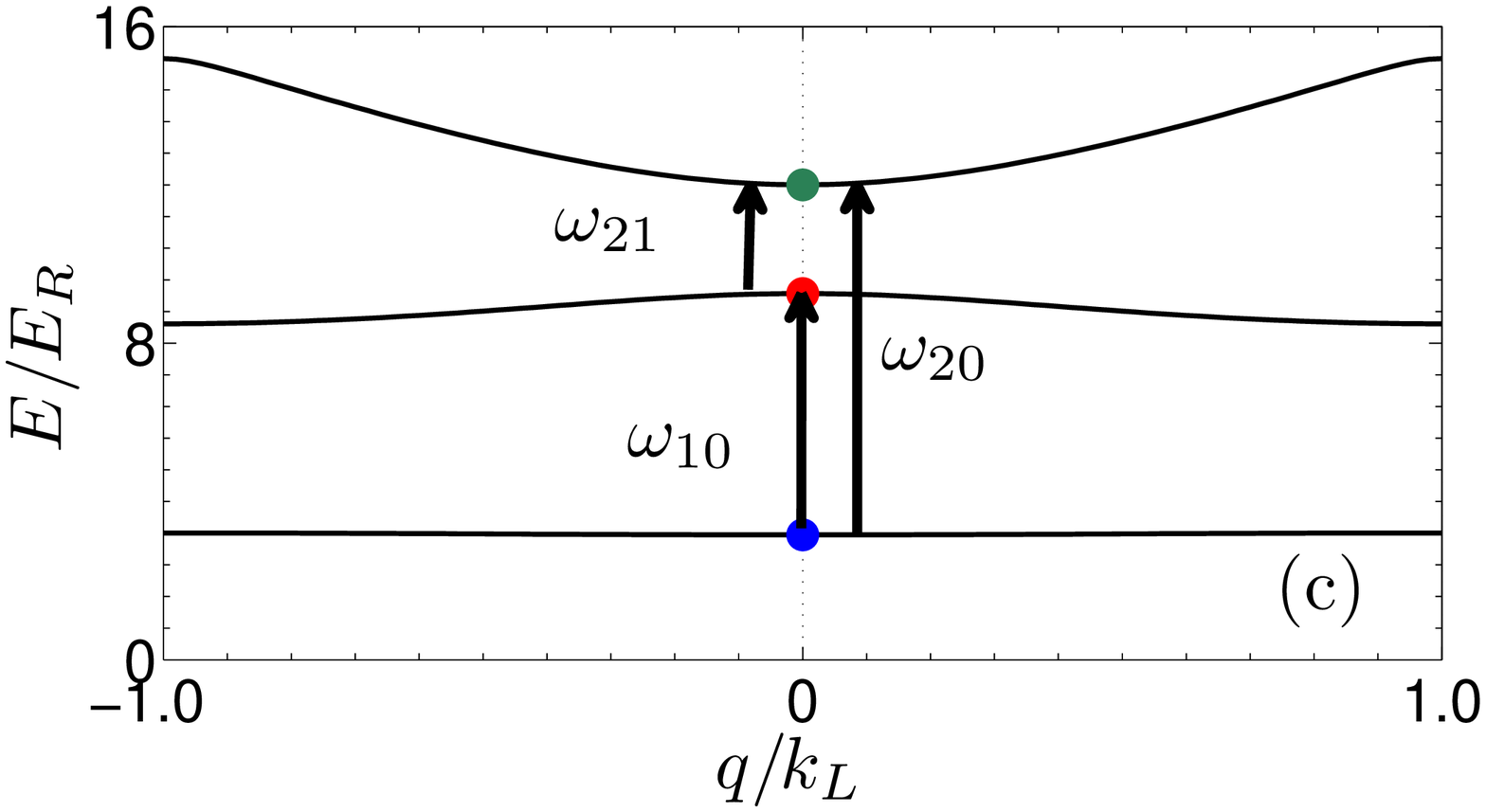}
\caption{(Color online)
From the TOF images, we can extract (a) the average momentum of the
cloud and (b) the relative occupation number of the $-1$ order
with $p = - 2 \hbar k_L$. By fitting the results with Eqs. (\ref{eqn:occupation})
and (\ref{eqn-average p}) (red solid), we can extract the frequencies
between multi energy bands, as schematically drawn in (c).
The results correspond to the case of $V = 12 E_R$, and all
date points are taken as average of more than three experimental trails.}
\end{figure}

Given the expressions of Eqs (\ref{eqn-average p}) and
(\ref{eqn:occupation}), we can extract the three frequencies from
evolution of the average momentum $\langle p \rangle$ [Fig. 3(a)],
and the relative number of particles in $p = - 2 \hbar k_L$ state
[Fig. 3(b)] via a fitting process, as depicted with solid (red)
curves. One should notice that since the frequency $\omega_{21}$ is
relatively small, it would require longer evolution time $\tau$ to
be precisely determined. Considering the typical range of $\tau
\lesssim 250 \mu$s in our experimental scheme, a more practical
strategy is to fit the other two frequencies $\omega_{10}$ and
$\omega_{20}$ first, and rely on the relation $\omega_{21} =
\omega_{20} - \omega_{10}$ to get $\omega_{21}$.


\begin{figure}[t]
\begin{center}
\includegraphics[height=6.8cm]{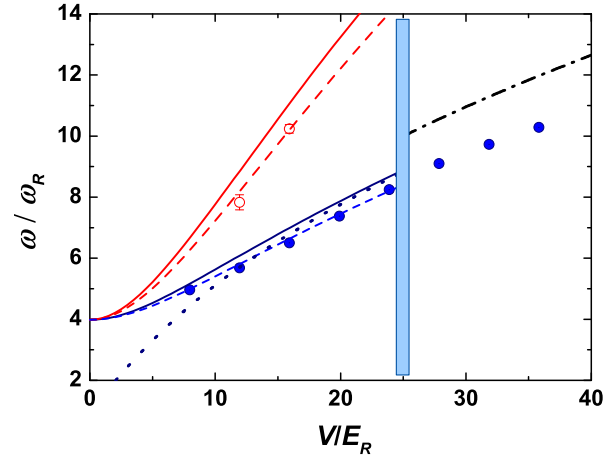}
\caption{(Color online) Oscillating frequencies $\omega_{10}$ (solid
dots) and $\omega_{20}$ (hollow dots) in the unit of recoiled
frequency $\omega_R=E_R/\hbar$. In the BEC regime with $V \le 25
E_{\rm R}$ (left), experimental results are systematically below the
bare lattice calculation (solid), where interaction effect is not
taken into account. To incorporate the interaction effect, an
effective potential theory (dashed) and a tight-binding model
(dotted) are considered. In the MI/normal regime with $V > 25 E_{\rm
R}$ (right), the oscillating frequency $\omega_{10}$ is compared
with the trapping frequency of the lattice potential
(dashed-dotted), where we use $a_s = 4.7$ nm, and $n_0 = 10^{14}$
cm$^{-3}$.}
\end{center}
\end{figure}

In Fig. 4, we show the results of oscillating frequencies of $\omega_{10}$
and $\omega_{20}$ for various optical lattice depths ranging from 8 to 37 $E_R$.
Notice that the frequency $\omega_{20}$ can only be obtained for lattice depth
$V \le 16 E_R$, as the strong lattice potential prohibits significant excitation
to the $(n=2)$ band.
For shallow lattices with $V \lesssim 25 E_R$, the system is in the BEC
regime presenting clear interference pattern in the TOF images,
and the discussion above remains valid. If we neglect the interaction
effect between atoms, the band structure can be calculated based on the bare optical
lattice potential, and the mutual gaps between different bands are indicated as
solid lines. As one would expect, this estimation is systematically above our
experimental observation, since the repulsive interaction tends to
reduce the lattice depth to a lower effective value.

One theory to incorporate the interaction effect is discussed in
Ref.~\cite{choi-99}, where Choi and Niu analyzes the interaction
effect for the bottom ($n=0$) band and gives an approximate
expression for the effective potential $V_{\rm eff} = V /(1+ 4C)$ by
assuming a nearly uniform condensate distribution. Here, $C = \pi
n_0 a_s /k_L^2$ is the dimensionless interaction strength and $n_0$
the {\it average} three-dimensional condensate density. By using the
effective potential, the oscillating frequencies can be extracted as
mutual gaps between different bands, as illustrated in Fig. 4
(dashed lines). Another method to capture the interaction effect is
to use a tight-binding model~\cite{kramer-03}, which is more
reasonable for deeper lattices since the Wannier functions for all
bands are well localized within one lattice site. Thus, we can
calculate the bottom and first exited band structures via a
variational method by neglecting higher order overlaps between
sites, and extract the frequency $\omega_{10}$ (dotted line in Fig.
4). We emphasize that neither of the theories above has included
finite temperature nor inhomogeneity effect, and their comparison to
our experimental data can only be understood to a qualitative level.


Up to now, we have restricted our discussion to the shallow lattices regime where
the system is in the BEC phase. By further increasing the lattice height over
25 $E_R$, the strong local confinement would
eventually break the global phase coherence between different sites, and the
system leave the BEC regime. In this case, since the recoil energy from
diffracted photons $E_B= 4 E_R$ is much smaller
than the confining potential, atoms that are kicked by the Bragg pulse are mainly
localized within one lattice site and swing to and fro around the site center.
If we neglect the interaction effect, the oscillating frequency is thus the trapping
frequency of the lattice potential, as shown in Fig. 4 (dashed-dotted line).
Notice that this bare lattice estimation is significantly above the experimental
measurements. A detailed description of the interaction effect would require
a solution of the hydrodynamic equation, which is beyond the scope
of the present manuscript.

\section{Semiclassical theory for the cooperative scattering process}
\label{sec:theory}

\begin{figure}[t]
\centering
\includegraphics[width=9.0cm]{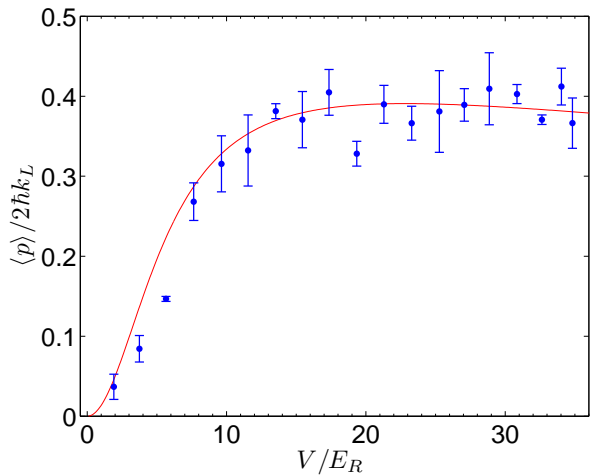}
\caption{(Color online) Average momentum of the cloud from TOF images upon
released from the trap {\it right} after the Bragg pulse of 5 $\mu$s.
Each data point is an average of more than three experimental trials.
This result can be well described by a semiclassical theory (solid, see text),
with fitting coupling constant $g = 0.9$ MHz.}
\end{figure}

Finally, we notice that the oscillating frequencies between
different energy bands can be identified only for fairly deep
optical lattices with $V \ge 8 E_{\rm R}$. For shallower lattices,
we could not observe significant oscillation. This can be easily
understood by reminding that the incident photons are coherently
scattered and amplified by the atomic grating, which becomes more
apparent in deeper lattice potentials. This scattering processes can
be described by using a semiclassical theory~\cite{ma-06}. We start
with the Maxwell-Schr{\"o}dinger equation, which describes the
evolution of atomic wave function $\Psi(x,t)$ and the positive and
negative frequency components of the classical electric field ${\bf
E^{(\pm)}}$:
\begin{eqnarray}
i\hbar \frac{\partial}{\partial t}\Psi(x,t)
&=&
\hat{H}\Psi+\frac{({\bf d}\cdot{\bf E^{(-)}})({\bf d}\cdot{\bf E^{(+)}})}{\hbar \delta}\Psi(x,t),
\label{MS-Equation1}  \\
\frac{\partial^2 {\bf E^{(\pm)}}}{\partial t^2}
&=&
c^2\nabla^2 {\bf E^{(\pm)}}-\frac{1}{\varepsilon_0}
\frac{\partial^2 }{\partial t^2}{\bf P^{(\pm)}}.
\label{MS-Equation2}
\end{eqnarray}
where $\hat{H}=-\frac{\hbar^2}{2M}\nabla^2+V(x)$ the single atom
Hamiltonian, $\delta$ the detuning of the laser from the electronic
transition, ${\bf d}$ the atomic dipole moment, and ${\bf
P^{(\pm)}}=-\frac{|d|^2|\Psi|^2}{\hbar \delta}{\bf E^{(\pm)}}$ the
polarization. By decomposing the atomic wave and electric field as
$\Psi=\sum_\ell \Psi_\ell e^{i\cdot 2\ell k_L x}$, and ${\bf
E^{(\pm)}}=\varepsilon_p^{\pm}e^{-i(\omega_p-k_L x-k_z
z)}+\varepsilon_s^{\pm}e^{-i(\omega_s+k_L x-k_z z)}$, we can get:
\begin{eqnarray}
&&\frac{\partial}{\partial t}\Psi_{\ell}
=
\frac{-i |d|^2}{\hbar^2\delta}
\left(
|\varepsilon_p^+ |^2\Psi_{\ell}
+ \varepsilon_p^+ \varepsilon_s^-\Psi_{\ell-1}
+ \varepsilon_s^+ \varepsilon_p^- \Psi_{\ell+1}
\right),
\nonumber \\
&&
\label{MS-Equation3} \\
&&\frac{\partial\varepsilon_s^+}{\partial t}
+ c \frac{\partial\varepsilon_s^+}{\partial x_s}
=
\frac{-i|d|^2\omega_s}{2\varepsilon_0\hbar\delta}
\Big(
\sum_{\ell} |\Psi_{\ell}|^2 \varepsilon_s^+
+
\sum_{\ell} \Psi_{\ell} \Psi_{\ell+1}^{\ast} \varepsilon_p^+
\Big),
\nonumber \\
&&
\label{MS-Equation4}
\end{eqnarray}
where $\varepsilon_p$ and $\varepsilon_s$ are the amplitude of the
pumping light and the scattered light, $\omega_p$ and $\omega_s$ are
the corresponding frequencies, $k_z$ is the wave vector
perpendicular to the lattice light, $x_s$ is the coordinate along
the direction of the scattered beam. The second term of Equation
(\ref{MS-Equation3})'s right side represents that an atom at
$\Psi_{\ell-1}$ absorb a photo from the pumping Bragg beam and emit
another into the scattered beam. During this process it receives
recoiled momentum of $2\hbar k_L$ and is transmitted into
$\Psi_{\ell}$ mode. On the other hand, the third term shows an atom
from $\Psi_{\ell+1}$ mode absorb a photo from the scattered beam and
emit one into the pumping beam and is transferred backward into
$\Psi_{\ell}$ mode. Equation (\ref{MS-Equation4}) shows the
evolution and propagation of the scattered beam. Note that it is the
atomic grating $\sum_{\ell}\Psi_{\ell}\Psi_{\ell+1}^\ast$ that
initially transfers photons into the scattered beam from the pumping
light.

Assuming that the amplitude of light field varies slowly, and
neglecting the spatial distribution of $\Psi_{\ell}$ and
$\varepsilon_s^{\pm}$~\cite{PRA06}, we get:
\begin{eqnarray}
\frac{\partial \tilde{c}_{\pm1}}{\partial t} &=&
\pm g(\tilde{c}_{\mp 1}^* \tilde{c}_0 + \tilde{c}_0^* \tilde{c}_{\pm 1} ) \tilde{c}_0,
\label{EF1}  \\
\frac{\partial \tilde{c}_0}{\partial t} &=&
g ( |\tilde{c}_{-1}|^2 - |\tilde{c}_1 |^2 ) \tilde{c}_0,
\label{EF2}
\end{eqnarray}
where $\tilde{c}_{\ell}(t)=\langle 2\ell \hbar k_L|\Psi(t)\rangle$ ,
the amplitude of $\ell^{\rm th}$ momentum mode, can be solved
numerically from the above equations, together with the initial
conditions $\tilde{c}_{\ell}(0)=c_{n=0}^{\ell}$.
$g=|d_{12}|^4|\varepsilon_p|^2 \omega_s nL / (2\epsilon_0 c \hbar^3
\delta^2)$ represent the coupling of the light with BEC of length
$L$. Using the relation $\tilde{c_{\ell}}=\sum_{n}c_n^{\ell}a_n$ we
can determine the coefficients $a_n$ and their evolution, hence
obtain the oscillation process discussed above. In Fig. 5, we show
the average momentum $\langle \hat{p}\rangle=2\hbar
k_L\sum_{\ell}\ell\cdot|\tilde{c}_{\ell}|^2$ of the cloud right
after a Bragg pulse of $5 \mu$s duration, and compare with the
semiclassical prediction. For shallow lattices, the momentum change
is relatively small such that the oscillatory behavior upon time
evolution is blurred. However, for deeper lattice potentials,
significant amount of momentum can be transferred to atoms via this
coherent scattering process.

\section{conclusion}
\label{sec:conclusion}

In summary, an experimental investigation of multi-band energy
spectrum for a bosonic system loaded in a 1D optical lattice is
demonstrated. A coherent Bragg light scattering process to excite
atoms to higher energy bands with zero pseudo-momentum transfer, and
oscillatory behavior of average momentum upon holding the sample of
atoms in the optical lattice for various evolution times, are first
reported. We extract the oscillating frequencies, which correspond
to the energy gap between different bands, for a wide range of
lattice potential depth. Throughout the whole range of lattice
depth, we observe the effect of repulsive interaction between
particles, which tends to reduce the lattice height to a lower
effective value. Our method is a coherent process where photons are
scattered and amplified by the atomic density grating, while the
previously reported Bragg spectroscopy requires a high coherence of
atoms.

\acknowledgments

This work is supported by the National Fundamental
Research Program of China under Grant No.
2011CB921501, the National Natural Science Foundation
of China under Grant No. 61027016, No.61078026,
No.10874008 and No.10934010. WZ would like to thank RUC for support (10XNF033, 10XNL016).

\end{document}